%% file: velocity_defects_revised_final.tex
\documentclass[11pt]{article}
\usepackage{fancyhdr}
\usepackage{isomath}
\usepackage{amsmath}
\usepackage{amsbsy}
\usepackage{amssymb}
\usepackage{amscd}
\usepackage{amsfonts}
\usepackage{graphicx,color}
\usepackage{verbatim}
\usepackage{euscript}
\usepackage{alltt}
\usepackage{stmaryrd}
\usepackage{subfigure}
\usepackage{url}

\input{temp-definitions}

\usepackage[margin=1in]{geometry}

\date{}
\begin{document}
\title{Some preliminary observations on a defect Navier-Stokes system\footnote{\bf To appear in Comptes Rendus - M\'ecanique.}}
\author{Amit Acharya\thanks{Department of Civil \& Environmental Engineering, and Center for Nonlinear Analysis, Carnegie Mellon University, Pittsburgh, PA 15213, email: acharyaamit@cmu.edu.} \and Roger Fosdick\thanks{Department of Aerospace Engineering \& Mechanics, University of Minnesota, Minneapolis, MN 55455, email: fosdick@aem.umn.edu.}
}
\maketitle

\begin{abstract}
\noindent Some implications of the simplest accounting of defects of compatibility in the velocity field on the structure of the classical Navier-Stokes equations are explored, leading to connections between classical elasticity, the elastic theory of defects, plasticity theory, and classical fluid mechanics.
\end{abstract}

\section{Introduction}
The objective of this brief note is to explore connections between the violation of the St.-Venant compatibility conditions in the theory of linear elasticity leading to the elastic theory of defects \cite{kroner1981continuum}, and its counterpart expression of incompatibility in classical fluid mechanics embodied in the Navier-Stokes equations. A modified `defect Navier-Stokes' model is postulated and some preliminary observations on its implications are made. A main purpose is to stimulate  a cooperative  exchange between fluid mechanists who are interested in high Reynolds number flows and solid mechanists who are engaged in defect theory related to dislocation--disclination dynamics.

\section{Kinematics of defects in the velocity field}
Consider the situation when the spatial velocity gradient field in a flowing fluid becomes singular on a surface of discontinuity, when viewed from a macroscopic length scale. Denote the \emph{`singular'} part of the velocity gradient by $L^{(s)}$. Then, one can define the \emph{`regular'} part as
\begin{equation}\label{eqn:Lr}
\supr{L} := \nabla \, v - \sups{L},
\end{equation}
where $v$ is the spatial velocity field of the fluid particles (the symbol $\nabla$ is meant as a distributional gradient). Thus, $L^{(r)}$ is an integrable field, containing, roughly speaking, at most bounded discontinuities. Similarly, now consider the gradient of the regular part of the velocity gradient; this may as well contain a singular part, $K^{(s)}$, and we again define the regular part of $\nabla L^{(r)}$ as
\begin{equation}\label{eqn:Yr}
Y^{(r)} := \nabla L^{(r)} - K^{(s)} = \nabla^2 \, v - \left( \nabla L^{(s)} + K^{(s)} \right).
\end{equation}
Thus, $\nabla L^{(s)} + K^{(s)}$ may be considered as the singular part of the second-gradient of the velocity field.
We note that
\begin{equation}\label{eqn:curl_r_s}
\begin{split}
curl \, L^{(r)} & = - curl \, L^{(s)},\\
curl \, Y^{(r)} & = - curl \, K^{(s)}.
\end{split}
\end{equation}
(all tensors will be expressed w.r.t a fixed orthonormal basis of a rectangular Cartesian coordinate system and all spatial derivatives are w.r.t the coordinates of the said system parametrizing the current configuration of the body; a $curl$ of a tensor amounts to a `row-wise' $curl$ based on the last index of the matrix of the tensor - this has an invariant meaning). In the theories we envisage, we expect the `singular parts' to also be integrable functions, at least, so partial differential equations, in some appropriate weak form, has a chance of being formulated for their evolution. Physically, this implies that we are adopting a microscopic point of view that resolves, with additional integrable fields, these macroscopically singular concentrations, the latter assumed to be manifestations of viewing, from a distance (or over a macrosocpic scale), localized smooth concentrations on a microscopic scale.  This is expected to be organized by the modified structure of the governing equations of the theory and its constitutive equations. With this understanding, the regular parts of fields are not expected to have any `large' concentrations even on `thin' sets (of full measure), at most containing bounded discontinuities.

In order to make a connection with Weingarten's theorem \cite{wein, delph_wein, volt, delph_volt, acharya2019_wein_volt}, we will consider the restricted case where $K^{(s)}$, a third-order tensor, is skew in its first two indices, which can also be expressed in terms of a second order tensor $\omega^{(s)}$ as 
\begin{equation}\label{eqn:omega_k}
K^{(s)} = X \inprod{1} \omega^{(s)}, \  K^{(s)}_{ijk} = e_{ijs}\omega^{(s)}_{sk}.
\end{equation}
Then,
\[
e_{rmk}\, \sups{K}_{ijk,m} = e_{ijs} e_{rmk}\, \sups{\omega}_{sk,m},
\]
and we define the velocity-disclination density or \emph{v-disclination density} as
\begin{equation}\label{eqn:visclination}
\theta = - curl \, \sups{\omega}, \qquad \theta_{sr} = - e_{rmk} \sups{\omega}_{sk,m}.
\end{equation}
We also define the velocity-dislocation density or \emph{v-dislocation density} as
\begin{equation}\label{eqn:vislocation}
\alpha := - \supr{Y} \inprod{2} X = curl \, \supr{L} + \sups{K} \inprod{2} X = - curl \, \sups{L} + \sups{K} \inprod{2} X.
\end{equation}
Since
\[
\sups{K}_{ijk}e_{jkl} = -e_{jis}e_{jkl} \sups{\omega}_{sk} = - \left(\delta_{ik}\delta_{sl} - \delta_{il} \delta_{sk} \right) \sups{\omega}_{sk} = - \left( \sups{\omega}_{li} - \sups{\omega}_{kk} \delta_{il} \right),
\]
we note that
\begin{equation}\label{eqn:ahat}
\widehat{\alpha} := \alpha + \left(\sups{\omega}\right)^T - tr\left(\sups{\omega}\right) I = -curl \, \sups{L} 
\end{equation}
is a $curl$ of a field.  

Physically, the support of the $curl$ of  the singular part of the gradient of a field, when concentrated around a curve, represents the termination of a surface of discontinuity of the field in question.

Since both $\widehat{\alpha}$ and $\theta$ are defined through the $curl$ operation, they admit natural balance statements related to topological charge of lines/cylinders of support where these fields are concentrated \cite{ach_frac}. These cylinders cannot end in the body so that they are closed loops or run from boundary to boundary of the body (mathematically, a $curl$ is divergence-free). The conservation law for topological charge carried by these fields can be expressed, in the Eulerian setting, in the form \cite[Sec. 3]{acharya2007jump}
\begin{subequations}\label{eqn:conserv_law}
\begin{align}
\label{eqn:conserv_law_1}
\parderiv{\theta}{t} & = - curl \left(\theta \times \left(v + V^{(\theta)}\right) \right),\\
\label{eqn:conserv_law_2}
\parderiv{\widehat{\alpha}}{t} & = - curl \left( \widehat{\alpha} \times \left(v + V^{(\widehat{\alpha})} \right) \right),
\end{align}
\end{subequations}
where $V^{(\theta)}$ and $V^{(\widehat{\alpha})}$ are the spatial velocities of the $v$-disclination and $v$-dislocation defect fields, respectively, relative to the fluid particles. These are to be constitutively specified.

To see the connection of the kinematic structure above to the St.-Venant compatibility condition of linear elasticity theory, we have from \eqref{eqn:ahat}
\begin{equation}\label{eqn:Stvenant}
\begin{split}
& e_{rqi} \left( \alpha^T \right)_{li,q} + e_{rqi} \sups{\omega}_{li,q} - e_{rql} \left( \sups{\omega}_{kk,q} \right)  = - e_{rqi} e_{lkj}\sups{L}_{ij,kq} = - e_{rqi} e_{lkj} \sups{D}_{ij,kq}\\
\implies & \left( curl\, \widehat{\alpha}^T \right)_{sym} = \left( curl\, \alpha^T \right)_{sym} - \theta_{sym}  = - curl \left(curl\,\sups{D}\right)^T =:  - inc\, \sups{D},
\end{split}
\end{equation}
noting that $(inc\, A)_{sym} = 0$, for $A$ skew-symmetric and $\sups{D} := \sups{L}_{sym}$. The symbol $inc$ is the St.-Venant compatibility operator, and we see that in the presence of $v$-dislocations and $v$-disclinations in the body, $\sups{D}$ cannot generally be represented as the symmetrized gradient of a velocity vector field, and neither can $\supr{D}$ from \eqref{eqn:Lr} and linearity of the operator $inc$.
\section{A minimally modified Navier-Stokes system and discussion}
Assume that $V^{(\theta)} = V^{(\widehat{\alpha})} = 0$, i.e., the $v$-disclination and $v$-dislocation defects move with the fluid velocity, and consider the system
\begin{subequations}\label{eqn:N-S_system}
\begin{align}
\label{eqn:mod_NS_2} 
\parderiv{\widehat{\alpha}}{t} & = - curl \left( \widehat{\alpha} \times  v \right),\\
\label{eqn:mod_NS_5}
- inc\, \sups{D} & = \left( curl\, \widehat{\alpha}^T \right)_{sym},\\
\label{eqn:mod_NS_6}
\parderiv{\rho v}{t} + div (\rho v \otimes v) & = div \left( -p I + \mathbb{C} \left(D - \sups{D} \right)\right),\\\label{eqn:mod_NS_7}
\left( \mathbb{C} \left( D - \sups{D} \right)\right) \inprod{2} D & \geq 0,\\
\label{eqn:mod_NS_8}
\frac{d \rho}{dt} + \rho \, div \, v & = \parderiv{\rho}{t} + div\, (\rho \,v)  = 0,
\end{align}
\end{subequations}
where $D = (\nabla v)_{sym}$, and $\frac{d \rho}{dt}$ is the material time derivative of the mass density $\rho$. For a  compressible fluid $\mathbb{C}$ is a constant, positive-definite, isotropic, fourth-order tensor, characterized by two scalar constants, the bulk and shear viscosities $(\mathbb{C} = \lambda I \otimes I + 2 \mu \mathbb{I})$, where $I$ is the second-order identity and $\mathbb{I}$ is the fourth-order identity tensor on the space of symmetric tensors, and within a purely mechanics setting the scalar $p$ is then the constitutively determined `thermodynamic pressure,' typically a function of the density $\rho$. For an incompressible fluid it follows that $\frac{d \rho}{dt} = div \, v = 0$, and $\mathbb{C}$ is then characterized by a single constant $(\mathbb{C} = 2 \mu \mathbb{I}, \mu > 0)$ and the field $p$ is now not determined constitutively, but instead through the field equations \eqref{eqn:N-S_system} and prescribed boundary and initial conditions. Equations \eqref{eqn:mod_NS_6}-\eqref{eqn:mod_NS_7}-\eqref{eqn:mod_NS_8} represent the balance of linear momentum, the second law of thermodynamics within a purely mechanics setting, and the balance of mass, respectively. 

In the system \eqref{eqn:N-S_system}, \eqref{eqn:conserv_law_1} is not included since \eqref{eqn:Stvenant} shows that it is sufficient to consider the evolution of only $\widehat{\alpha}$ to understand the effect of the defect fields $\widehat{\alpha}, \theta$ on the solution for the velocity field (and mass density) through \eqref{eqn:mod_NS_6}-\eqref{eqn:mod_NS_7}-\eqref{eqn:mod_NS_8}. In general, the evolution of $\widehat{\alpha}$ and $\theta$ together define the evolution of the $v$-dislocation density, $\alpha$, from \eqref{eqn:visclination}-\eqref{eqn:ahat}, a special simplifying case being the assumption that $\theta = 0$.

We make the following observations:
\begin{enumerate}
\item  We first consider the restricted case wherein we assume that $div \, \left( \mathbb{C} \sups{D} \right) = 0$. 

In this case, the model may be interpreted as follows: when the fluid velocity field is such that the $D$ field is regular, i.e., $\sups{D} = 0$, then we have the standard N-S system. When $\sups{D} \neq 0$ in the domain, we consider two further cases:
\begin{itemize}
\item Assume $\sups{K} = 0$, $\sups{\omega} = 0$ (which implies $\theta = 0$) i.e., the gradient of the regular part of the velocity gradient is regular. 

Then, every singular solution of the classical N-S system \eqref{eqn:mod_NS_6} and  \eqref{eqn:mod_NS_8} now needs to satisfy an additional nonlocal differential inequality \eqref{eqn:mod_NS_7} arising from $\sups{D}$ being a functional of the velocity field obtained by solving the system \eqref{eqn:mod_NS_5} 
and \eqref{eqn:mod_NS_2} with $\widehat{\alpha}$ replaced by $\alpha$. 

There are obvious `gauge invariances' here, e.g., $\sups{D}$ is unique up to additions of symmetrized gradients of vector fields \eqref{eqn:mod_NS_5}, and considering the set of all possible $\sups{D}$ fields corresponding to each velocity field that solves  \eqref{eqn:mod_NS_6} and \eqref{eqn:mod_NS_8}, \eqref{eqn:mod_NS_7} appears to be a statement of convexity of the set (with $D$ being a point on its boundary). If boundary (and interior) conditions are invoked to supply uniqueness of solutions to \eqref{eqn:mod_NS_5}\footnote{For instance, as subsequently mentioned in the text, if $D^{(s)}$ is \emph{defined} as $D^{(s)} := inc \, \chi$ where $\chi$ is a solenoidal symmetric tensor field, then $\chi$ satisfies the inhomogeneous biharmonic equation, a fourth-order elliptic equation whose conditions for well-posedness are well-understood (see, e.g., \cite{rektorys1977variational}). Such boundary conditions may be imposed, justified by the requirement that $D^{(s)}$ should vanish when the rhs of \eqref{eqn:mod_NS_5} vanishes. },
then \eqref{eqn:mod_NS_7} is a nonlocal inequality constraint.

The similarity of notions of yield functions in rigid plasticity is to be noted.
\item Suppose now $\sups{K} \neq0$. The situation is mathematically as in the previous case but the physical meaning of the variable $\widehat{\alpha}$, and consequently the set of admissible $D^{(s)}$ fields for a given $v$ field is different, and this difference arises from the differing initial conditions (and inflow boundary conditions) that are imposed on $\widehat{\alpha}$ in \eqref{eqn:mod_NS_2} and that are compatible with the conditions $\sups{K} = 0$ and $\sups{K} \neq 0$. For example, suppose, for a certain 2-d flow, that both the $v$-dislocation and the $v$-disclination fields, $\alpha$ and $\theta$, are not identically $0$, but vanish outside a simply connected subregion $\Omega$ that is strictly contained within the (simply connected) 2-d flow domain  with  the $v$-disclination strength $\int_\Omega \theta \, \nu \,da \neq 0$, where $\nu$ is the outward unit normal to $\Omega$. Then the field $\widehat\alpha$ cannot vanish outside any subregion of the flow domain that contains $\Omega$ because, by \eqref{eqn:ahat}, $\widehat{\alpha} = 0$ would require the field $\omega^{(s)}$ (and, consequently $K^{(s)}$)  to vanish there as well. But, because of \eqref{eqn:visclination} (and an application of Stokes theorem), this contradicts the hypothesis that the $v$-disclination strength does not vanish. 

\end{itemize}
As an aside, following the procedures of the stress-function approach to linear elasticity \cite{kroner1981continuum}, if we assume the restricted ansatz $\sups{D} = inc \, \chi$ with $\chi$ symmetric and $div \, \chi = 0$, then $div\, \sups{D} = 0$, and  $ inc \,\sups{D} = \Delta^2 \chi$, where $\Delta^2$ is the biharmonic operator, $\Delta^2(\cdot) = (\cdot)_{,jjkk}$, thus giving familiar structure to the solutions of \eqref{eqn:mod_NS_5} \footnote{If the fluid is incompressible, then in the restricted case under consideration we have   $div\, D^{(s)} = 0$. Thus, if the fluid domain is nonparaphractic (i.e., no embedded holes) we may apply the representation $D^{(s)} = inc \,\chi$ with $div\, \chi = 0$ without loss of completeness. In \cite{Ach-Fos}, we restrict the general theory presented here to the study of velocity dislocations in incompressible fluid dynamics, and, assuming $div \,D^{(s)} = 0$, we give an elementary example of the theory for the case of plane Poiseuille flow of an incompressible fluid between parallel plates. The fluid is driven by a steadily increasing pressure gradient. At a specified Reynolds number the classical steady laminar flow is assumed to switch to an incompatible steady flow, which we determine. A description of the nonsteady transition that takes place between these two steady flows is left for a future study.}. 
\item Of course, through \eqref{eqn:mod_NS_6}, it is possible to consider the more general situation of $div \left( \mathbb{C} \sups{D} \right) \neq 0$, in which case there is a direct coupling between the velocity and $\sups{D}$ fields. Note that due to space-time filtering in the usual way (see, e.g., \cite{babic1997average}), a different contribution than the usual Reynolds stress tensor will appear in the averaged total stress tensor from the average of the $\sups{D}$ field. To see this, consider the case of an incompressible fluid and the following averaging operator:

For any microscopic field $q$ (viewed as a function of the variables $(x',t')$), the meso/macroscopic space-time averaged field $\bar{q}$ is given by
\begin{equation}\label{eqn:avg_oper}
\overline{q}(x,t)  := \frac{1}{\int_{\Omega(x)} \int_{I(t)} w(x-x', t-t')dx' dt'  } {\int_{\Lambda} \int_{B} w(x-x', t-t')\, q(x',t') \, dx' dt'},
\end{equation}
where $B$ is the flow domain and $\Lambda$ is a sufficiently large interval of time. $\Omega(x)$ is a bounded region within the domain around the point $x$ with linear dimension of the spatial resolution of the model to be developed, and $I(t)$ is a  bounded interval contained in $\Lambda$. The weighting function $w$ is non-dimensional and non-negative scalar valued, and assumed to be  smooth in the variables $x, x', t,$ and $ t'$. For fixed $x$ and $t$, $w$ has support in $\Omega(x) \times I(t)$ when viewed as a function of $x'$ and $t'$. Considering \eqref{eqn:N-S_system} as a system for microscopic fields (with differential operators in terms of the microscopic space-time variables) and applying the averaging operator \eqref{eqn:avg_oper} to each equation of the system, noting that for an incompressible fluid $\rho$ is a constant and that $\partial_{x_i} w = - \partial_{x'_i} w, \partial_t w = - \partial_{t'} w$, one obtains (up to a `layer' around $\partial B$) the \emph{averaged} equations
\begin{subequations}\label{eqn:avg_N-S_system}
\begin{align}
& \parderiv{\overline{\widehat{\alpha}}}{t}  = - curl \left( \overline{\widehat{\alpha}} \times  \overline{v}  + \overline{{\sf F}^{\,\widehat{\alpha} \times v}} \right), \notag\\
& {\sf F}^{\,\widehat{\alpha} \times v} (x,x',t,t')  := \widehat{\alpha}\left(x',t'\right) \times v\left(x',t'\right) - \overline{\widehat{\alpha}}(x,t) \times \overline{v}(x,t), \notag\\
& \overline{{\sf F}^{\,\widehat{\alpha} \times v}}  = \overline{\widehat{\alpha} \times v} - \overline{\widehat{\alpha}} \times \overline{v}, \ \mbox{ equivalently} \  \ \overline{{\sf F}^{\,\widehat{\alpha} \times v}} (x,t)  = \overline{\left(\widehat{\alpha} - \overline{\widehat{\alpha}}(x,t)\right) \times \left(v - \overline{v}(x,t)\right) }\,(x,t),  \notag\\
& - inc\, \overline{\sups{D}}  = \left( curl\, \overline{\widehat{\alpha}}^T \right)_{sym}, \notag\\
\label{eqn:avg_linmom}
& \parderiv{\rho \overline{v}}{t} + div (\rho \overline{v} \otimes \overline{v})  = div \left( -\overline{p} I + \mathbb{C} \left(\overline{D} - \overline{\sups{D}}\right) - \overline{{\sf F}^{\,\rho v \otimes v}} \right), \\
&{\sf F}^{\,\rho v \otimes v} (x,x',t,t') := \rho \, \left( v(x',t') \otimes v(x',t') - \overline{v}(x,t) \otimes \overline{v} (x,t) \right), \notag\\
& \overline{{\sf F}^{\,\rho v \otimes v}} = \rho\, \overline{v \otimes v} - \rho \, \overline{v} \otimes \overline{v}, \ \mbox{ equivalently} \  \  \overline{{\sf F}^{\,\rho v \otimes v}}(x,t) = \rho\, \overline{(v - \overline{v}(x,t)) \otimes (v - \overline{v}(x,t))}\, (x,t), \notag\\
\label{eqn:avg_sec_law}
&0 \leq  \left( \mathbb{C} \left( \overline{D} - \overline{\sups{D}} \right)\right) \inprod{2} \overline{D} + \overline{{\sf F}^{\mathbb{C}(D - D^{(s)}) \inprod{2} D} }, \\
 &{\sf F}^{\left(\mathbb{C}\left( D - D^{(s)} \right)\right) \inprod{2} D} \left(x,x',t,t'\right)  := \left( \mathbb{C}\left( D\left(x',t'\right) - D^{(s)}\left(x',t'\right) \right)\right) \inprod{2} D\left(x',t'\right)  \notag\\
& \qquad \qquad \qquad \qquad \qquad \qquad \quad - \left( \mathbb{C}\left( \overline{D} - \overline{D^{(s)}} \right)\left(x,t\right)\right) \inprod{2} \overline{D}\left(x,t\right), \notag\\ 
& div \, \overline{v}  = 0,  \notag
\end{align}
\end{subequations}
where all differential operators are w.r.t. the coordinates $(x,t)$, and we note that it is only under very special circumstances that averages of products are equal to products of averages. The `independent' fields of the averaged model above are $( \overline{\widehat{\alpha}}, \overline{v})$. While the fields $\overline{\sf F^{(\cdot)}}$ are well-defined in terms of the fields of the `microscopic' system \eqref{eqn:N-S_system}, they cannot, in general, be expressed as functionals of the independent fields of the averaged model, such expressions being referred to as `closure' relations. It is clear, given the microscopic model \eqref{eqn:N-S_system}, that evolution equations for the $\overline{\sf F^{(\cdot)}}$ fields can as well be written down, in which case they become (the $\overline{\sf F^{(\cdot)}}$s) additional independent fields and they, in turn, give rise to further higher-order `fluctuations,' and without closure, the cycle continues. In this manner an infinite hierarchy of equations arises but, of course, it is the closure equations for the fluctuations up to any given/inferred retained level that are of paramount fundamental importance as well as the determination of at what level, if at all, the microscopic dynamics \eqref{eqn:N-S_system} implies a closed averaged model.
\item From work in \cite{garg2015study}, nucleation of the defect field $\widehat{\alpha}$ with null initial conditions can be expected due to the coupling of $\widehat{\alpha}$ to the velocity field $v$ in \eqref{eqn:conserv_law}.
\item 

A different modification is to have $V^{(\theta)}, V^{(\widehat{\alpha})}$ constitutively specified, increasing the ad-hoc nature of the model. An interesting and strong difference between defect dynamics in elasticity and the present fluid model is that in solids, the defect fields in the \emph{inverse deformation gradient} have an effect upon the stored energy and the Second Law of Thermodynamics constrains the constitutive specification of the defect velocity fields. In the viscous fluid case being discussed, the defects in the velocity field do not affect the stored energy, and the Second Law, as it stands, does not appear to provide any direct guidance, at least of the type available in solids, on the constitutive specification of the defect velocities. Thus, for this modification, the Second Law, as it stands in classical fluid dynamics, may not be adequate, and a critical examination of the physical admissibility of additional energy related contributions due to the existence of  defects in fluids is warranted.

\item An interesting question is whether or not $v$-dislocations and $v$-disclinations can be experimentally observed in flow visualizations. For example, one situation which may be amenable to identification of singularities through analysis of flow visualization is when these defects are isolated, i.e., consider the case of a 2-d flow in which the fields $\alpha$ and $\theta$ are localized and vanish outside a subregion $\Omega$ of the 2-d flow domain. The line integral of $L^{(r)}$ on a closed loop in `good' far-field fluid can presumably be evaluated and when $\widehat{\alpha} = 0$, we see that \eqref{eqn:ahat} and \eqref{eqn:curl_r_s} require this integral to vanish  regardless of the contour. When $Y^{(r)} = \nabla L^{(r)}$, which by \eqref{eqn:Yr} requires $K^{(s)} = 0$, i.e., the `regular flow curvature' field does not have a singular part, then $\sups{\omega}$ vanishes and, because of \eqref{eqn:curl_r_s}, \eqref{eqn:ahat}, and Stokes theorem we see that  the closed-loop integral of $L^{(r)}$ is a constant on all contours that encircle $\Omega$ and given by $\int_\Omega \alpha  \nu \, da$, where $\nu$ is the outward unit normal to $\Omega$. Thus the non-zero constancy of the closed-loop integral on the said contours when $\int_\Omega \alpha  \nu \, da \neq 0$ is a necessary condition for the the absence of `flow-curvature' singularities, i.e., the absence of $\sups{K}$ and $curl \,Y^{(r)} = 0$. The experimental identification of such isolated `singularities' can also help in testing the assumption that these defects potentially move with the fluid velocity.

\item A primary remaining question is if the structure of this model can be useful in the study of the N-S equations (especially as applied to high Reynolds number flows), since for such flows it poses additional constraints on solutions of the classical N-S system, alternative to those that are traditional within the theory of turbulence.

We note that the classical N-S system and its various idealizations model the observed behavior of real fluids remarkably well. Robust computational capabilities for approximating the solutions of such systems are well developed and much used for practical purposes. It is perhaps fair to state that the main challenges that remain in understanding the N-S equations relate to mathematical questions of well-posedness of solutions, and the theoretical and practical questions of developing models for, and computing, the effective response of turbulent fluids at larger length scales than the viscous length scale approaching atomic dimensions. In what follows, we provide some (speculative) remarks about the possibilities of our proposed model in aiding the study of some such issues.

Here, we consider various idealizations of the compressible and incompressible fluid within a thermodynamic setting. Since our efforts relate to improving models of discontinuities in classical continuum mechanics, it is worthwhile to first recall and comment upon the well-known expression of mass balance at a surface of discontinuity: $\llbracket \rho (v \cdot n  - u_n) \rrbracket = 0$ \cite{truesdell1960classical,landau_lif}, where $n$ represents the unit normal to the surface of discontinuity, $u_n$ is the normal velocity of the surface of discontinuity, and $\llbracket (\cdot ) \rrbracket$ represents the `jump' of $(\cdot )$ across the surface. This makes it clear that for a (homogeneous) incompressible fluid, shocks, i.e., $\llbracket v \cdot n \rrbracket \neq 0$, are not allowed. Discontinuities in the tangential velocity (slip surfaces or vortex sheets) are allowed in both compressible and incompressible fluids. As is well-known, flat tangential discontinuities in the inviscid, incompressible case are unstable (the Kelvin-Helmholtz instability, \cite[Sec. 29, 84]{landau_lif}), giving way to a turbulent region, with similar phenomena arising in the viscous and compressible \cite[Sec. 84]{landau_lif},\cite{fjordholm2017construction} cases. We remark that in the setting of this note, the singular part of the velocity gradient $\sups{L}$ is a field that can be non-vanishing in `thin' sets of full measure. The sets can be time-dependent and `curved' in space, with their `terminations,' i.e. the locations where $\widehat{\alpha} = -curl\, \sups{L} \neq 0$, serving as the `core' of the vortices that we believe are so abundantly seen in well-developed turbulent regions following such instabilities.

The explicit thermomechanical governing equations for the various flow idealizations are:
\begin{itemize} 
\item {\bf Case I - the compressible, viscous fluid} 
\begin{subequations}\label{eqn:comp_viscous}
\begin{align}
\label{eqn:mod_cv_3}
& \parderiv{\rho}{t} + div\, (\rho \,v) = 0, \\
\label{eqn:mod_cv_4}
& \parderiv{}{t}(\rho E) + div (\rho Ev) = div (T v - q); \qquad E:= \varepsilon(\rho, \eta) + \half \rho \, v\cdot v; \qquad q := - \kappa \nabla \vartheta,\\
& T:= - \hat{p}(\rho, \eta) I + \lambda\, tr \left(D - \sups{D}\right)I + 2\mu \left(D - \sups{D} \right); \qquad \hat{p}(\rho, \eta): = \rho^2 \parderiv{\varepsilon}{\rho} (\rho, \eta), \notag\\
\label{eqn:mod_cv_5}
& \parderiv{}{t}(\rho v) + div (\rho v \otimes v) = div\, T, \\
\label{eqn:mod_cv_6}
& \left( \lambda\,  tr\left(D - \sups{D}\right)I + 2\mu \left(D - \sups{D} \right) \right) \inprod{2} D  \geq 0; \qquad 0< \vartheta := \parderiv{\varepsilon}{\eta} (\rho, \eta),\\
\label{eqn:mod_cv_1} 
& \parderiv{\widehat{\alpha}}{t} = - curl \left( \widehat{\alpha} \times  v \right),\\
\label{eqn:mod_cv_2}
& - inc\, \sups{D} = \left( curl\, \widehat{\alpha}^T \right)_{sym},
\end{align}
\end{subequations}
where balance of energy (the First Law of thermodynamics) has also been included with $\varepsilon$ being the internal energy density per unit mass, $\vartheta$ is the absolute temperature, $\eta$ the entropy density per unit mass, $q$ is the heat flux vector, and $\kappa$ is the diffusivity. The inequalities $\lambda + \frac{2}{3} \mu > 0$, $\mu > 0$, and $\kappa > 0$ hold, and we note from \eqref{eqn:mod_cv_6} that $\eta$ may be expresed as a function $\vartheta$ and $\rho$.

Here and in the following Cases along with their various assumptions on $\lambda$ and $\mu$, the ansatz 
\[
div \,( \mathbb{C} \sups{D}) = 0  \implies div\, \left( \lambda \,tr \left(\sups{D}\right)I + 2\mu \sups{D} \right) = 0,
\]
may be a helpful simplifying feature. \emph{In what follows, this ansatz will be assumed in all Cases discussed.}

We note that system \eqref{eqn:mod_cv_3} and \eqref{eqn:mod_cv_5} are exactly their classical counterparts, but the thermodynamic statements \eqref{eqn:mod_cv_4}, \eqref{eqn:mod_cv_6} include the effects of $\sups{D}$. Existence of weak solutions in space dimension 3 for the classical N-S equations for this case has been proven in \cite{bresch2007existence} but uniqueness of solutions, or the lack of it, has not yet been proven. Our modified system for $(\rho, v, \eta)$ is not the same as the classical case as just outlined, and it remains to be seen whether the full system \eqref{eqn:comp_viscous} can aid in the mathematical study of the compressible, linearly viscous fluid for high Reynolds number flows.
\item {\bf Case II - the compressible, inviscid fluid under adiabatic conditions, i.e., the compressible Euler equations}

This idealization applies particularly to gases for which it is reasonable to assume the bulk viscosity $\lambda + \frac{2}{3} \mu = 0$ and $q = 0$. 

In order to develop the governing system for this case, we first consider this idealization as a special case of the compressible viscous fluid, i.e., Case I, under the ansatz that $\lambda + \frac{2}{3} \mu = 0$ and $q = 0$. Noting that $\mu > 0$ in Case I, this ansatz implies that \emph{\eqref{eqn:mod_cv_6}$_1$ is equivalent to 
\begin{equation}\label{eqn:diss_ind_mu}
\left(D - \sups{D} \right)_{dev} \inprod{2} D_{dev} \geq 0,
\end{equation}
a statement independent of $\mu$}, where $( \cdot)_{dev}$ refers to the deviatoric part of a second-order tensor.  Hence, under the ansatz $\lambda + \frac{2}{3} \mu = 0$ and $q = 0$, system \eqref{eqn:comp_viscous} can be written as stated with \eqref{eqn:mod_cv_6} replaced by  $\left(D - \sups{D} \right)_{dev} \inprod{2} D_{dev} \geq 0$. Now, taking the formal limit  $\mu \to 0$ in this specialization of \eqref{eqn:comp_viscous}, we obtain the system
\begin{subequations}\label{eqn:comp_inviscid}
\begin{align}
\label{eqn:mod_cinv_3}
& \parderiv{\rho}{t} + div\, (\rho \,v) = 0, \\
\label{eqn:mod_cinv_4}
& \parderiv{}{t} (\rho E) + div (\rho Ev) = div (-\hat{p}(\rho, \eta) v); \qquad E:= \varepsilon(\rho, \eta) + \half \rho \, v\cdot v, \\
\label{eqn:mod_cinv_5}
& \parderiv{}{t}(\rho v) + div (\rho v \otimes v) = div (- \hat{p}(\rho, \eta) ); \qquad \hat{p}(\rho, \eta): = \rho^2 \parderiv{\varepsilon}{\rho} (\rho, \eta), \\
\label{eqn:mod_cinv_6}
& \left(D - \sups{D} \right)_{dev} \inprod{2} D_{dev} \geq 0,\\
\label{eqn:mod_cinv_1} 
& \parderiv{\widehat{\alpha}}{t} = - curl \left( \widehat{\alpha} \times  v \right),\\
\label{eqn:mod_cinv_2}
& - inc\, \sups{D} = \left( curl\, \widehat{\alpha}^T \right)_{sym}.
\end{align}
\end{subequations}
The equations \eqref{eqn:mod_cinv_3},\eqref{eqn:mod_cinv_4},\eqref{eqn:mod_cinv_5} form the classical system for $(\rho, v, \eta)$ for this case. In the classical setting, shocks and tangential discontinuities are known to arise and persist in this case, for special problems. Within this setting, the existence of global weak solutions for representative problems containing a wide class of initial data  is not known (with very practically demonstrable consequences \cite{fjordholm2017construction}), let alone uniqueness; an excellent discussion is provided in \cite{ fjordholm2017construction} including computational aspects. As discussed previously, the equations \eqref{eqn:mod_cinv_6}, \eqref{eqn:mod_cinv_1}, \eqref{eqn:mod_cinv_2} provide additional constraints on weak `solutions' of the classical theory, were they to exist, and may be expected to affect the uniqueness of such solutions. These equations may also be expected to constrain the entropy-measure-valued solutions and statistical solutions that have been defined \cite{fjordholm2017construction,fjordholm2017statistical}, following a distinguished history of work by L. Tartar, R. J. DiPerna, and C. Foias.
\item {\bf Case III - the incompressible, viscous fluid under isothermal conditions}

The governing equations in this case become
\begin{subequations}\label{eqn:incomp_viscous}
\begin{align}
\label{eqn:mod_incv_3}
& div\, v = 0, \\
\label{eqn:mod_incv_5}
& \parderiv{}{t}(\rho v) + div (\rho v \otimes v) = \frac{1}{\rho} div \left(- p I + 
2 \mu \left( D - \sups{D} \right)\right), \\
\label{eqn:mod_incv_6}
& \left(D - \sups{D} \right)_{dev} \inprod{2} D_{dev} \geq 0,\\
\label{eqn:mod_incv_1} 
& \parderiv{\widehat{\alpha}}{t} = - curl \left( \widehat{\alpha} \times  v \right),\\
\label{eqn:mod_incv_2}
& - inc\, \sups{D} = \left( curl\, \widehat{\alpha}^T \right)_{sym},
\end{align}
\end{subequations}
where $\rho > 0$ is a given constant mass density, $\mu >0$ holds, and $p$ now is a field not determined by a constitutive assumption but determined as part of the solution to a particular problem that is posed. 

Again, \eqref{eqn:mod_incv_3},\eqref{eqn:mod_incv_5} are exactly the classical equations for $(p, v)$, and \eqref{eqn:mod_incv_6},\eqref{eqn:mod_incv_1},\eqref{eqn:mod_incv_2} provide further constraints on solutions of \eqref{eqn:mod_incv_3},\eqref{eqn:mod_incv_5} due to the incompatibility of the fluid motion. The status of the mathematical study of \eqref{eqn:mod_incv_3},\eqref{eqn:mod_incv_5} may be found in \cite{bardos2013mathematics}. Global weak  (Leray-Hopf) solutions in 3-d are known to exist, but the status of their uniqueness is not known \cite{guillod2017numerical}(some very `rough' weak solutions have recently been shown to be non-unique \cite{buckmaster2019nonuniqueness}).
\item {\bf Case IV - the incompressible, inviscid fluid under isothermal conditions, i.e., the incompressible Euler equations}

Viewing this case as a formal limit of $\mu \to 0$ in \eqref{eqn:incomp_viscous}, the governing equations in this case become
\begin{subequations}\label{eqn:incomp_inviscid}
\begin{align}
\label{eqn:mod_incinv_3}
& div\, v = 0, \\
\label{eqn:mod_incinv_5}
& \parderiv{}{t}(\rho v) + div (\rho v \otimes v) = \frac{1}{\rho} div \left(- p I \right), \\
\label{eqn:mod_incinv_6}
& \left(D - \sups{D} \right)_{dev} \inprod{2} D_{dev} \geq 0,\\
\label{eqn:mod_incinv_1} 
& \parderiv{\widehat{\alpha}}{t} = - curl \left( \widehat{\alpha} \times  v \right),\\
\label{eqn:mod_incinv_2}
& - inc\, \sups{D} = \left( curl\, \widehat{\alpha}^T \right)_{sym},
\end{align}
\end{subequations}
where where $\rho > 0$ is a given constant mass density, and $p$ is again a non-constitutive fundamental field to be determined. 

The constraints provided by the proposed model \eqref{eqn:incomp_inviscid} to the classical system consisting of \eqref{eqn:mod_incinv_3}, \eqref{eqn:mod_incinv_5} are the same as in Case III. For the classical system, non-unique, \emph{dissipative}, H\"{o}lder continuous weak solutions have recently been constructed, as described in the excellent review \cite{delellis_szekel}, rigorously establishing Onsager's conjecture on anomalous dissipation in the incompressible Euler equations (i.e., there exists (continuous) velocity fields whose total kinetic energy strictly decreases in time if the velocity field satisfies the condition $|v(x,t) - v(y,t)| < C |x - y|^h \ \forall x,y,t$ ($x,y$ are locations, $t$ is time) with  $ h < \frac{1}{3}$, where $C >0 $ is a constant independent of $x,y,t$), as well as far-reaching connections to the geometric isometric embedding problem.
\end{itemize}

Finally, we note that in Cases I and III above, and without the restriction $div\, \left( \mathbb{C} \sups{D} \right) = 0$, the equations as averaged according to the scheme in \eqref{eqn:avg_oper}-\eqref{eqn:avg_N-S_system} have an added contribution to the averaged stress tensor arising from $\overline{\sups{D}}$ that is expected to affect the coarse-scale modeling of turbulence by our model. (We note that the averaging operator  \eqref{eqn:avg_oper} can easily be adjusted to not include any averaging in time if desired, e.g., as is done in Large Edddy Simulations (LES) for modeling intermittency.) In all Cases, the averages of the `dissipation', \eqref{eqn:mod_cv_6}, \eqref{eqn:mod_cinv_6}, \eqref{eqn:mod_incv_6}, and \eqref{eqn:mod_incinv_6} have added contributions in comparison to their classical counterparts.

\end{enumerate}
\bibliographystyle{alpha}\bibliography{velocity_defects_ref}

\end{document}

%% file: temp-definitions.tex
\usepackage{amsmath}
\usepackage{amsbsy}
\usepackage{amssymb}
\usepackage{amscd}
\usepackage{amsfonts}

\newcommand{\beq}{\begin{equation}}
\newcommand{\eeq}{\end{equation}}
\newcommand{\beqs}{\begin{eqnarray}}
\newcommand{\eeqs}{\end{eqnarray}}
\newcommand{\beql}{\begin{equation} \label}
\newcommand{\half}{\frac{1}{2}}

\newcommand{\parderiv}[2]{\frac{\partial #1}{\partial #2}}

\newcommand{\inprod}[1]{\cdot_{#1}}

\newcommand{\sups}[1]{#1^{(s)}}
\newcommand{\supr}[1]{#1^{(r)}}